\documentclass[traditabstract]{aa} 

\usepackage{graphicx}
\usepackage{txfonts}
\usepackage{url}
\usepackage{natbib}
\bibpunct{(}{)}{;}{a}{}{,} 
\usepackage[utf8]{inputenc}
\usepackage{subfigure}
\usepackage{verbatim} 
\begin{document}

   \title{Observational evidence for buffeting induced kink waves in solar magnetic elements}
   \author{M. Stangalini$^{1}$, G. Consolini$^{2}$,  F. Berrilli$^{3}$, P. De Michelis$^{4}$, R. Tozzi$^{4}$ }
   \institute{$^{1}$ INAF-Osservatorio Astronomico di Roma, 00040 Monte Porzio Catone (RM), Italy\\
   $^{2}$ INAF-Istituto di Astrofisica e Planetologia Spaziali, 00133 Roma, Italy\\
   $^{3}$ Universita degli Studi di Roma Tor Vergata, 00133 Roma, Italy \\
   $^{4}$ Istituto Nazionale di Geofisica e Vulcanologia, 00143 Roma, Italy \\
   \email{marco.stangalini@inaf.it}}

  \abstract 
{The role of diffuse photospheric magnetic elements in the energy budget of the upper layers of the Sun's atmosphere has been the recent subject of many studies. This was made possible by the availability of high temporal and spatial resolution observations of the solar photosphere, allowing large numbers of magnetic elements to be tracked to study their dynamics.  In this work we exploit a long temporal series of seeing-free magnetograms of the solar photosphere to study the effect of the turbulent convection in the excitation of kink oscillations in magnetic elements. We make use of the empirical mode decomposition technique (EMD) in order to study the transverse oscillations of several magnetic flux tubes. This technique permits the analysis of non-stationary time series like those associated to the horizontal velocities of these flux tubes which are continuously advected and dispersed by granular flows.\\
Our primary findings reveal the excitation of low frequency modes of kink oscillations, which are sub-harmonics of a fundamental mode with a $7.6 \pm 0.2$ minute periodicity. These results constitute a strong case for observational proof of the excitation of kink waves by the buffeting of the convection cells in the solar photosphere, and are discussed in light of their possible role in the energy budget of the upper Sun's atmosphere.}  

   \authorrunning{M. Stangalini}
	\titlerunning{Buffeting induced kink waves in magnetic elements}
\maketitle

\section{Introduction}
It is widely regarded that small magnetic elements in the solar atmosphere host a rich variety of MHD waves that may have an active role in the energy budget of the upper layers of the Sun's atmosphere. This possibility appears even more attractive since high resolution observations have shown that small magnetic elements with diameters $ \sim 100-150$ km cover a significant fraction of the solar photosphere \citep{2010ApJ...723L.164L, 2012A&A...539A...6B}, although it has been shown that magnetic flux emergence is not homogeneous \citep{2014A&A...561L...6S}.\\
After being generated and amplified, possibly by the action of small-scale dynamo (SSD) acting in the solar convection zone \citep{2011ApJ...737...52L, 2013A&A...555A..33B}, the small-scale magnetic fields are perturbed by the external photospheric plasma flows. Under this action the flux tubes are shaken and, at the same time, they are advected toward the boundaries of the supergranular cells \citep{2011ApJ...743..133A, 2012ApJ...758L..38O, 2012ApJ...759L..17L, 2013ApJ...770L..36G, 2014arXiv1405.0677G}, to form the so-called magnetic network \citep{1989ApJ...345.1060S}.  \citet{2012ApJ...758L..38O}, employing a long  time series of high resolution magnetograms ($8$ hours),  observed the small magnetic concentrations within a supergranule moving radially outward from the center toward the network, with a velocity aligned to the plasma flow.  In this regard, \citet{2014arXiv1405.0677G} observed a change of the diffusion regime of small magnetic elements with a reduction of the diffusivity in proximity of the magnetic network. This implies different dynamic regimes in supergranular cells.\\
At the same time many oscillations (kink, sausage, etc.) are expected in the flux tubes in response to the external plasma forcing \citep{lrsp-2005-3}. This scenario is supported by numerical simulations \citep{Khomenko2008, Fedun2011, 2012ApJ...755...18V, 2012A&A...538A..79N}, and observations 
\citep{1995A&A...304L...1V, 2009Jess,2011ApJ...730L..37M, 2012ApJ...744L...5J, 2012ApJ...746..183J, 2013A&A...549A.116J, 2013A&A...554A.115S}, as well as by several theoretical studies \citep[to name a few]{1978SoPh...56....5R, 1981A&A....98..155S, Edwin1983, Roberts1983, Musielak1989, 1998ApJ...495..468S, Hasan2003, Musielak2003a}.\\
\citet{2013A&A...554A.115S} observed that kink waves coexist with longitudinal compressive waves in the same flux tube at the same time, and are closely related, possibly through non-linear interactions. \\
Although solar convection is commonly believed to be the primary driver of waves in flux tubes, little is known about the excitation mechanisms of these waves.\\
\citet{Hasan2003}, \citet{Musielak2003, Musielak2003a}, and  \citet{2008ApJ...680.1542H} argued that the photospheric forcing of magnetic elements can be considered a viable mechanism through which energy is extracted from the photosphere and channelled to the upper layers of the atmosphere.  It has been shown, in fact,  that the rapid footpoint motion due to the turbulent granular buffeting can effectively excite kink waves that can propagate upward and couple with longitudinal waves \citep{1997ApJ...486L.145K, Hasan2003}.
\citet{2011ApJ...740L..40K}, by using high resolution observations and simulations of the solar photosphere, found that while the majority of small-scale magnetic elements have velocities between 0 and 1 km/s, $6\%$ of them have velocities in excess of 2 km/s. The same authors argued that this significant fraction of magnetic elements may play an important role in the energy budget of the upper layers of the solar atmosphere, as velocities larger than 2 km/s can effectively drive kink waves \citep{1993ApJ...413..811C}. These velocities are consistent with the results found by many authors \citep[see for example][to mention a few]{2006SoPh..237...13M, 2012MSAIS..19...93C, 2012ApJ...752...48C, 2013SoPh..tmp..276B, 2013A&A...554A.115S}. \\
\citet{2012ApJ...746..183J} detected slow upward propagating longitudinal waves in small magnetic elements, using high cadence broad-band 2D data, visible as periodic intensity fluctuations in the range $110-600$ s. This provided an observational proof of the propagation scenario.\\
\citet{2013ApJ...768...17M} have shown, by exploiting high spatial and temporal resolution observations, that torsional modes \citep{2007Sci...318.1572E} generated by vortices in the solar photosphere can effectively propagate to the chromosphere. \\ Very recently, \citet{2014ApJ...784...29M} have investigated the propagation of incompressive waves to the chromosphere, finding a good agreement between photospheric and chromospheric velocity power spectra at frequencies below $8 mHz$.\\ 
\citet{2013A&A...559A..88S} have observed the presence of high frequency peaks of up to $10-12$ mHz in the power spectra associated to the horizontal velocity of individual magnetic elements. These high frequency oscillations are well above the cutoff frequency expected for kink waves, which is always smaller than the acoustic cutoff at $\sim 5.3$ mHz \citep{1981A&A....98..155S}, therefore they are expected to propagate upwards. Although the physical properties of the magnetic elements analysed were quite similar, \citet{2013A&A...559A..88S} found that each individual magnetic element has its own signature in the power spectrum. This was interpreted by considering large scale physical parameters such as the connectivity of each magnetic element and the complexity of the magnetic carpet that make the power spectrum of the horizontal oscillations of a flux tube not only dictated by the local physical properties of the flux tube.\\
\citet{2013A&A...559A..88S} estimated the response function of the magnetic elements subject to the forcing of the photospheric plasma flows by comparing the median power spectrum of kink oscillations in magnetic elements with the spectrum of the horizontal velocity associated to the granulation found by \citet{2010ApJ...716L..19M}. This response revealed that magnetic elements are prone to oscillate at high frequency ($ \nu > 5-6$ mHz).\\
These results, as well as most of those present in the literature so far, were obtained by employing FFT-based techniques. Since the process associated to the horizontal swaying of the flux tubes is far from being stationary, the use of FFT-based techniques may pose some limitation. Indeed, the modal basis used for the decomposition of the signals in FFT analysis is not well suited for analysing non-stationary processes.\\
To overcome this shortcoming in this work we analyse the horizontal fluctuations of several long-lived photospheric flux tubes making use of empirical mode decomposition (EMD), a technique that is based on a modal basis that is defined on top of the signal itself \citep{huang1998empirical}. This allows us to highlight some peculiarities in flux tube oscillations as well as a method of providing direct observational proof of the non-linear interaction between the external plasma and the magnetic elements, which excite MHD waves.

   \begin{figure}[t]
   \centering
   \subfigure{\includegraphics[width=9.5cm, clip]{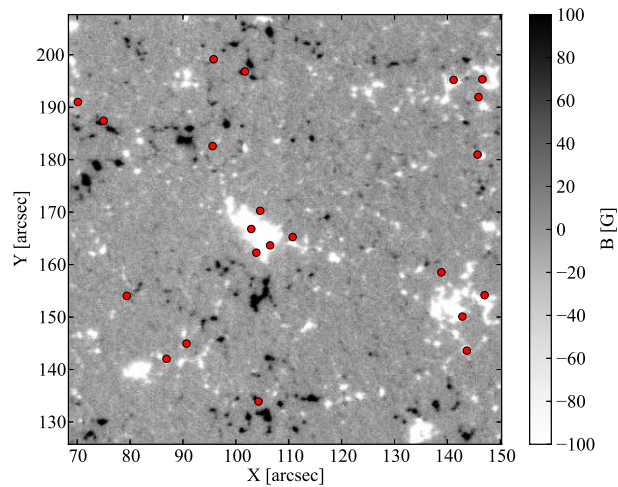}}
   \caption{Time averaged magnetogram obtained by averaging over the whole four-hours data set (saturated between $-100$ G and $100$ G). The red dots indicate the mean position of the longest-lived magnetic elements ($\geq 4800$ s) considered in the analysis.} 
    \label{map}
   \end{figure} 

\section{Data set and analysis}
The data set used in this work consists of a sequence of high spatial resolution magnetograms acquired by SOT/NFI, the narrow band imager on board Hinode satellite \citep{springerlink:10.1007/s11207-008-9174-z}. The magnetograms were obtained in the Na I $589.6$ nm spectral line from shutterless V and I Stokes filtergrams taken on 2008 August 18, close to disc centre. The cadence of the data is $30$ s. The diffraction limit in NFI filtergrams at $589.6$ nm is $0.24$ arcsec, while the pixel scale is set at $0.16$ arcsec. This results in a slight spatial downsampling in the focal plane, providing a $2$-pixel Nyquist spatial sampling of $0.32$ arcsec (or $\sim  230$ km on the solar photosphere).\\
The field-of-view (FoV) is approximately $80 \times 80$ arcsec, and the total duration of the time sequence is approximately $4$ hours. In Fig. \ref{map} the average magnetogram is shown. The FoV encompasses a few supergranules whose borders are highlighted by the network patches visible in the figure.\\
   \begin{figure}[t]
   \centering
   \subfigure{\includegraphics[width=9cm, clip]{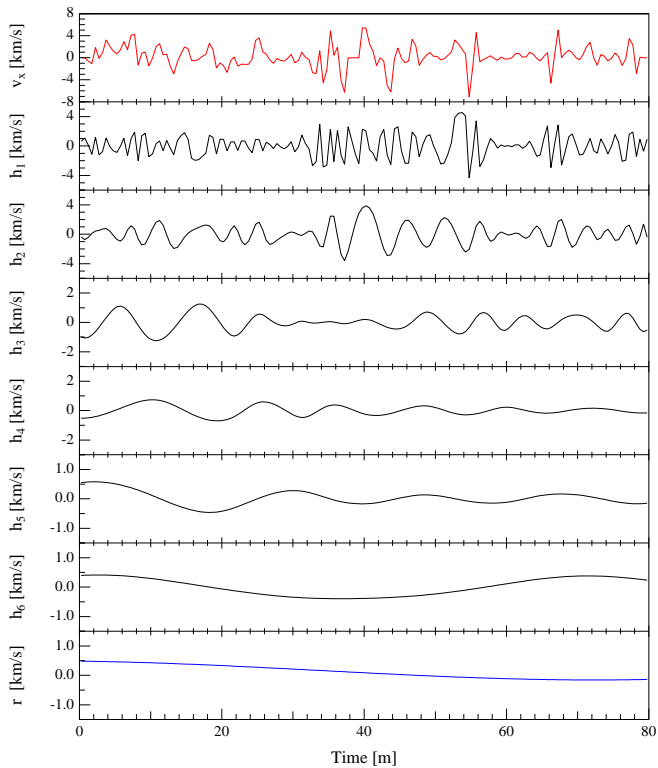}}
   \caption{Empirical mode decomposition of the horizontal velocity of a single magnetic element into six intrinsic mode functions ($h_{1}-h_{6}$), and a residue (r).}
    \label{emd}
   \end{figure} 
Along with the standard SOT/FG calibration procedure  (the IDL code \textit{prep-fg} available in the Hinode Solarsoft package), we also apply an additional calibration to limit the effects of jitter and tracking residuals. The visual inspection of the data in fact reveals a low frequency trend in the tracking and a few sudden shifts of the FoV which are not properly handled by the calibration code. Since we are mainly interested in the study of transverse oscillations of small magnetic elements, it is important to properly co-align the data set with sub-pixel accuracy. A registration procedure allowing sub-pixel accuracy must be applied. This procedure is based on FFT cross-correlation and utilizes the whole FoV to estimate any misalignment between two images. We apply FFT registration iteratively until the mean residuals are minimized. This happens in three registration iterations.\\
To study the dynamics of small magnetic features we apply the YAFTA (Yet another Feature Tracking Algorithm) tracking algorithm \citep{Welsch2003, 2007ApJ...666..576D}. This algorithm identifies and tracks magnetic pixels belonging to the same local maximum. To ensure the reliability of the results, three constraints are applied. Each magnetic feature must lie above a threshold and must have an area slightly larger than the full-width at half-maximum (FWHM) of the PSF (a $4 \times 4$ pixel box in our case) and a lifetime long enough to ensure a high spectral resolution. The threshold on the magnetic signal is chosen to be $2 \sigma$. Following \citet{2012SoPh..279..295L} we estimate the sigma of the magnetic signal by fitting a Gaussian to the low-field pixels with absolute value of the magnetic flux density below $200$ G. This resulted in a sigma  of $11.8$ G. Lastly, only the magnetic elements with a lifetime greater or equal to $4800$ s are considered in the analysis (22 elements in total). This sets the frequency resolution at $\sim 0.2 ~mHz$. The threshold on the lifetime (4800 s) was chosen as a tradeoff between the number of elements and the frequency resolution. \citet{2011SoPh..269...13T} and \citet{2013A&A...554A.115S} have extensively tested this tracking code on high resolution data. \\
In Fig. \ref{map} we show the average position of each magnetic element selected. 
From the position of each magnetic element given by the tracking algorithm the horizontal velocities are estimated. The position of the magnetic elements is estimated with sub-pixel accuracy as a flux-weighted average of the coordinates of the pixels constituting the magnetic element itself. The computation of the horizontal velocity using both the $x$ and $y$ components suffer from frequency doubling if the components themselves are oscillating around zero. This is a direct consequence of the definition of the total velocity where the squared value of the components is considered. In order to avoid this drawback in the spectral analysis we consider only the $x-$component of the horizontal velocity, hereafter $v_{h}$ for simplicity.\\

\subsection{Empirical mode decomposition}
The empirical mode decomposition (EMD) was originally introduced by \citet{huang1998empirical} as a precondition technique for the application of the Hilbert transform to a signal. The EMD moves from the assumption that most of the actual time series of natural systems are the result of the superposition of many different time scales, simultaneously active, and aims to decompose a signal into a set of intrinsic mode functions (IMFs), representing different oscillations at a local level. As opposed to the Fourier method, which decomposes a signal into a pre-fixed set of elementary functions (sine or cosine oscillations) on the basis of the energy content of each characteristic time scale and applies these to rigorously stationary time series, the characteristic time scales of the IMFs obtained by the EMD are based on the local distance between two successive extrema. In other words, an IMF represents a hidden oscillation mode, locally defined and thus not stationary. Indeed, an IMF is defined as a local mono-component mode, which satisfies the following conditions: \textit{i)} the number of extrema and zero-crossing should be equal or differ at the most by one, \textit{ii)} at any time the mean value of the upper and lower envelopes, defined from the local maxima and minima respectively, is zero. It is important to remark that EMD technique decomposes a signal without any \textit{a priori} assumption, providing a decomposition which is based on the data itself, i.e., \textit{a posteriori} decomposition.

The EMD algorithm was employed to extract the IMFs of a time series $x(t)$ consists of a sequence of steps that can be resumed as follows:
\begin{enumerate}
\item extract the local maxima and minima of the time series $x(t)$,
\item construct the upper ($e_u(t)$) and lower ($e_l(t)$) envelopes from local maxima and minima,
\item calculate the mean $m(t) = (e_u(t)+e_l(t))/2$,
\item define the new time series $h_i(t) = x(t)-m(t)$,
\item check if $h_i(t)$ satisfies the condition required to be an IMF, if not repeat the steps 1-5 on the time series $h_i(t)$
\item if $h_i(t)$ satisfies the condition to be an IMF, define a new signal $x^i(t) = x(t) - h_i(t)$ and start the procedure again to  iteratively define the other IMFs,
\item the iteration is stopped when the $x^k(t)$ has less than two extrema. In this case $x^k(t)$ is identified as the residue $r(t)$. 
\end{enumerate}
The steps from (1) to (5) constitutes the so-called \textit{sifting process}, which is stopped on the application of one of the stoppage criteria, proposed in the wide range of literature about this technique \citep{huang1998empirical, huang2003confidence, flandrin2004empirical,  huang2008amplitude}. The application of a stoppage criterion is required to validate that the IMFs have a physical meaning.  Here, we  apply the stoppage criterion used by \citet{rilling2003empirical}, fixing a maximal number of iterations of 300 steps and checking that the IMFs satisfy condition \textit{i)}. 
   \begin{figure}[t]
   \centering
   \subfigure{\includegraphics[width=8cm, clip]{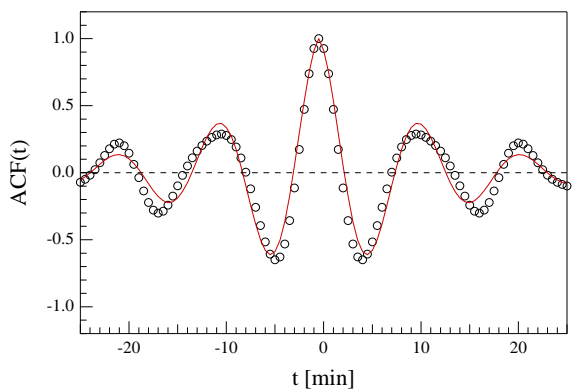}}
   \caption{Example of ACF fitting to estimate the main frequency of each IMF.}
   \label{acf}
   \end{figure} 
As a result of the EMD technique the original signal $x(t)$ is decomposed into a set of IMFs and a residue, so that we can write:
\begin{equation}
x(t)= \sum_{i=1}^n h_i(t)+r(t),
\end{equation}
where $\{h_i(t)\}$ is the set of IMFs and $r(t)$ is the residue. It is important to underline that the EMD method, being based on the extraction of the energy contained in the intrinsic time scales of the signal, is used to detect the rapid frequency changes that are typical features of non-stationary signals. Furthermore, as underlined in \citet{datig2004performance}, compared to the usual Fourier Transform, since the IMFs are not given in a closed analytical form, the mathematical orthogonality of the IMFs is only locally satisfied, instead of the  usual theoretical sense ($\langle h_i(t) | h_j(t) \rangle = \delta_{ij}$).\\
The EMD technique has already been used in many contexts of solar physics, heliospheric and magnetospheric physics such as, but not limited to, the study of coronal oscillations \citep{2004ApJ...614..435T}, neutrino flux modulation \citep{2010ApJ...709L...1V, 2012ApJ...749...27V}, cosmic rays modulation \citep{2014ApJ...781...71L}, and magnetospheric studies \citep{npg-19-667-2012}.

   \begin{figure}[t]
   \centering
   \subfigure{\includegraphics[width=8cm, clip]{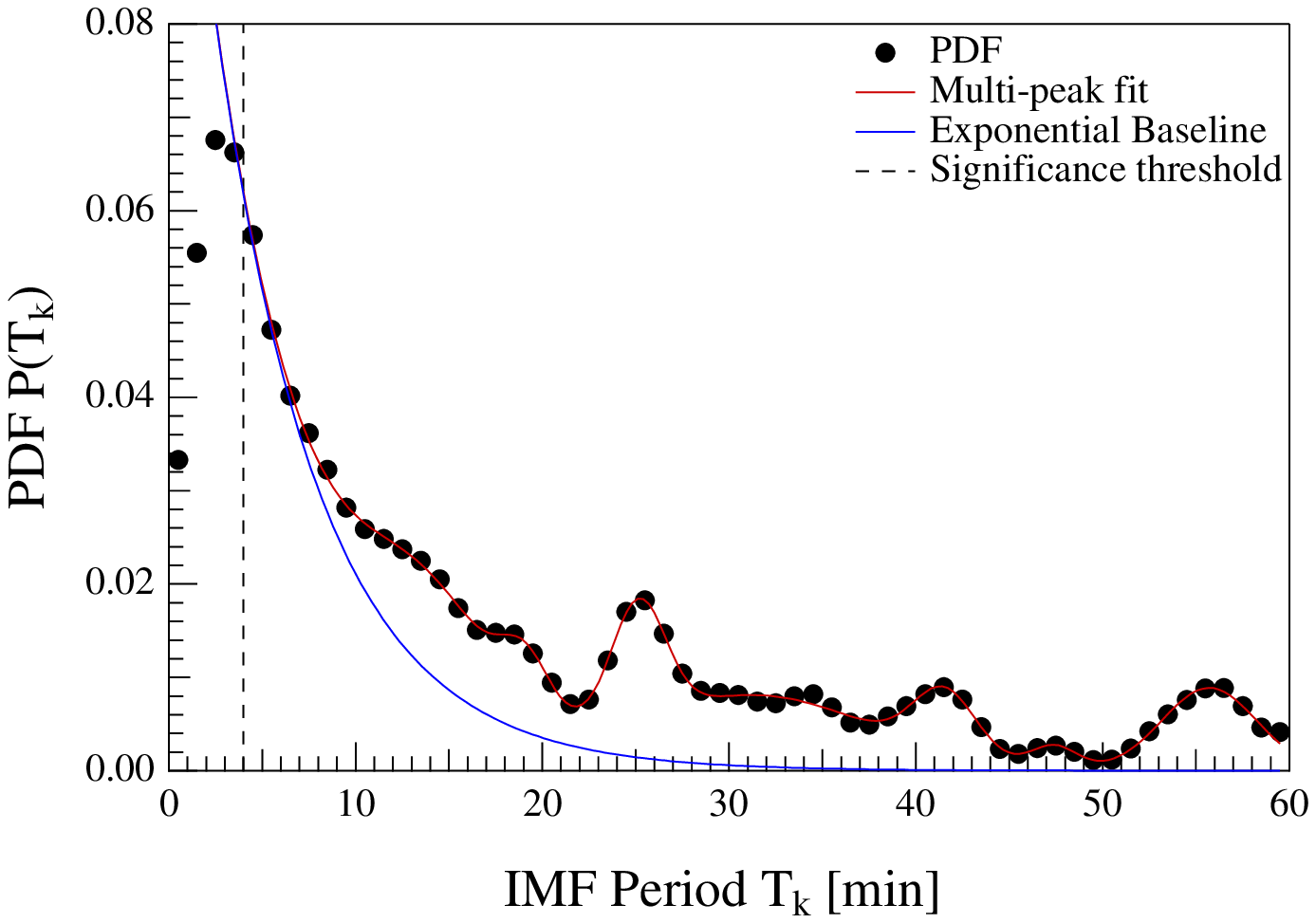}}
   \subfigure{\includegraphics[width=8cm, clip]{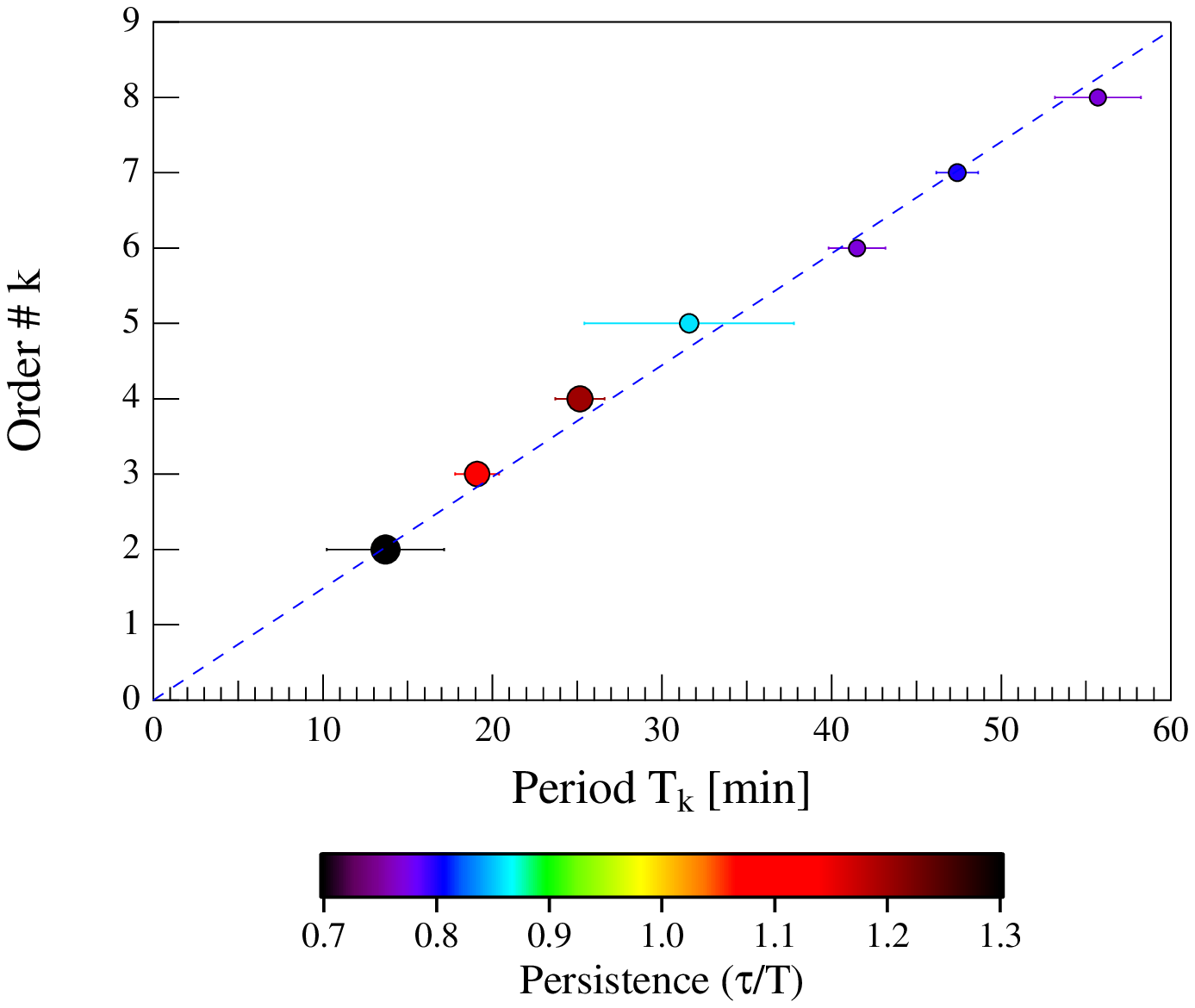}}

   \caption{Probability density function of the period (upper panel). Position of the peaks of the PDF as a function of the peak number (lower panel). }
   \label{harmonics}
   \end{figure}

\section{Results}
As described in the previous section, the EMD analysis provides a decomposition of the signals associated with the horizontal velocity of the selected magnetic elements in different intrinsic modes which are built on top of the time series itself. In our case, each $v_{h}$ time series is decomposed into six IMFs (see for example Fig. \ref{emd}).  In the upper panel of Fig. \ref{emd}  it is possible to see that the horizontal velocity of the flux tube shows strong fluctuations up to $\pm 4$ km/s. In addition, these strong fluctuations appear to be concentrated in short pulses in the time series, and the time series itself appears to be non-stationary.
For each IMF we estimated the main frequency through an analysis of the auto-correlation function (ACF, see Fig. \ref{acf}). This resulted in 132 frequency values in total (six values for each of the 22 magnetic elements). The ACF of each mode is fitted using an exponential function of the form:
\begin{equation}
ACF=A ~cos \frac{2 \pi x}{T_{k}} ~exp(-\frac{x}{\tau}),
\end{equation}
where $A$ is a constant term, $x$ represents the time lag, $T_{k}$ is the period which has to be estimated from the fit, and $\tau$ is a characteristic time scale (the characteristic length of autocorrelation of the signal). It is worth noting here that each IMF is not purely monochromatic. This is because unlike the FFT, where each mode decomposing the time series has a unique frequency associated to it, in EMD, each mode contains different frequencies, although very close to each other. For this reason by fitting the ACF one obtains the mean period of oscillation. This is also the reason why in Fig. \ref{acf} the fit of the ACF is not perfectly matching the ACF itself, but there are small differences.\\
In Fig. \ref{harmonics} (upper panel), we plot the probability density function (PDF) of all the mean periodicities found in each IMF of the decomposition and for each magnetic element. Since the total number of frequencies is limited (one value for each of the six IMFs and for each of the 22 magnetic elements), the PDF is obtained using the kernel method, as explained in \citet{kaiser2002information}. The kernel method is in fact an alternative to binning a distribution and is best suited for cases in which the number of samples is limited. This approach is mainly based on the assumption that the probability density is smooth enough that one can ignore all the structures below a certain kernel width. Instead of simply counting points within a certain bin, one can consider distance-dependent weights obtained by using a kernel function. In our case we made use of a Gaussian kernel function.  \\
The PDF increases progressively at small periods and presents a series of distinct bumps corresponding to the most common periodicities found in the signals. In order to check whether these peaks have a distinct spacing, an exponential function plus a series of gaussians peaks is fitted to the data. This allows us to estimate the position of each bump. The results of this analysis are shown in Fig. \ref{harmonics} (lower panel). In this figure the period of each peak is plotted against the order $k$ at which it is found. Here we start from the order $k=2$ by assuming that the order $k=1$ is hidden below the frequency resolution limit imposed by the length of the time series analyzed. This limit is represented by the dashed vertical line in the upper panel of Fig. \ref{harmonics}. It is easy to see that the spacing of the most common periodicities is not random.  The resulting scaling law is linear $T(n)=T_{0}n$ where $n$ represents the peak number and $T$ the period.\\
A linear fit to the data results in a slope $T_{0}$ corresponding to a period of $7.6 \pm 0.2$ min, where the error is obtained by a weighted linear regression. Here, the weights used are proportional to the reciprocal of the errors associated to the peak position as obtained by the multi-Gaussian nonlinear fit (Levenberg-Marquardt method) of the  PDF. This number is consistent with the time scale of granulation (between $5-10$ min). In addition, the scaling relation of the peaks is the signature of sub-harmonics where $T_{0}$ represents the fundamental period of the system. \\
Moreover, the EMD analysis shown in Fig. \ref{emd} also highlights the strong non-stationary nature of the signals under investigation which manifests rather clearly in the IMFs. All of them in fact display a frequency modulation and a rather common presence of wave trains with different characteristics (i.e. amplitude, modulation, and waiting time). \\
In the lower panel of Fig. \ref{harmonics}, we also show the persistence of each non-stationary oscillation compared to the period of the oscillation. This piece of information is encoded in the different colors and the different widths of the points of the plot itself. This persistence is given by the characteristic decorrelation scale, $\tau$,  estimated from the fit of the ACF mentioned above. Thus the persistence represents the temporal autocorrelation of the signal, that is, the duration of a wave train.  
What is clear is that the lower the period, the larger the persistence of a particular oscillation.

\section{Discussion}
Our analysis shows that at least the longest-lived magnetic elements in our data set ($22$ elements in total) present several periodicities which are consistent with sub-harmonics. The fundamental mode is estimated to be  $\sim 7.6 \pm 0.2$ min, a period which is close to the evolutionary time scale expected for granulation in the solar photosphere. This means that the buffeting action that the photospheric plasma exerts on the magnetic elements induces oscillations with a period which is consistent with the granular lifetime \citep[e.g.][]{1989ApJ...336..475T, 1999ApJ...515..441H, 2004A&A...428.1007D}. \\
This is evidence for buffeting-induced transverse oscillations in photospheric magnetic elements. These oscillations are strictly non-stationary and for this reason they are investigated employing a decomposition whose modal base can be obtained directly from the signals themselves. This allows us to overcome the well-known limitations of FFT-based techniques, when applied to non-stationary signals.\\
In addition, our results show the presence of a linear scaling of the periods of oscillations found the in the magnetic elements which is consistent with sub-harmonic oscillations.\\
It is worthwhile to briefly discuss here the mechanisms of excitation of sub-harmonics in a physical system. For a complete review of the topic we refer the reader to \citet{han2002nonlinear}.\\
Both sub-harmonics and super-harmonics can be either excited in linear or non-linear systems, although with some difference in the underlying mechanisms. \\
In linear systems, sub-harmonic or super-harmonic response can only be expected when exact relations between the natural and the forcing frequency are satisfied. That is, the forcing frequency must be an exact fraction or multiple of the natural frequency. 
On the other hand, in non-linear systems, the super-harmonic or sub-harmonic response can be excited by a range of forcing frequencies with values not necessarily multiples or fractions of the natural frequency. It has been shown, in fact, that in non-linear systems a response at a frequency other than the driving frequency is also observed \citep{1981PhRvL..47.1349L}. In particular sub-harmonics and super-harmonics in non-linear systems can be generated as a response to intense driving \citep{yen1971subharmonic}, although they have a different character. While super-harmonics are always generated in the presence of a non-linear driving source, sub-harmonics need a certain forcing strength in order to be excited \citep[see for example ][for a complete treatment of the topic]{stern1965theory}. Moreover, the presence of period-doubling cascades, like that observed in our results, can be seen as the signature of chaotic excitations in non-linear systems \citep[][to mention a few]{2009arXiv0910.3570S, 2010arXiv1002.3363S}. \\
\citet{2010ApJ...716L..19M}, using a local spatial cross-correlation technique, studied the horizontal oscillations of the velocity of the photospheric plasma, which results in a broad power spectrum. For this reason, it is difficult to distinguish between a linear and non-linear response of the flux tubes since in the linear regime the exact relation between the natural and the forcing frequency can be easily satisfied in the presence of a broad spectrum.\\
However, our results constitute a strong observational proof of the excitation of kink waves as a result of granular buffeting.\\
The flux tubes in the solar photosphere are embedded in a high-$\beta$ plasma whose forcing may induce both forced and free oscillations in the flux tube itself. \citet{2013A&A...559A..88S} found, studying a large number of photospheric magnetic concentrations, a small peak in the range $3-4$ mHz of the kink oscillation spectrum, although the large error bars associated did not allowed them to identify that peak with a high confidence level.\\ 
Our results extend the results of \citet{2013A&A...559A..88S} by analysing the power spectral density of kink oscillations in photospheric magnetic elements with the very large frequency resolution provided by the long duration of the time series associated to the elements selected ($80$ min). This allowed us to explore also the very low frequency part of the power spectrum ($\nu < 4$ mHz down to $0.2~mHz$) of kink oscillations.\\
Apart from this, our results also indicate that most of the energy is contained in lower frequency modes ($P> 25$ min), since the persistence of these modes is larger than the modes at higher frequency. \\
It is a well-known result that the cutoff frequency of kink waves is always smaller than its acoustic counterpart \citep{1981A&A....98..155S}. This is given by:
\begin{equation}
\omega_{kink}= \dfrac{\omega_{ac}}{2 \gamma (2 \beta +1)} ,
\end{equation}
where $\omega_{ac}$ represents the acoustic cutoff which is $\simeq 5.3$ mHz, and $\beta$ is the gas to magnetic pressure ratio, and $\gamma$ the adiabatic index. Under typical conditions the kink cutoff frequency in a flux tube is at least two times smaller than what is estimated. Of course it also depends on the inclination angle through $\omega_{ac}$ \citep{2006ApJ...647L..77M}. For this reason it is difficult to estimate its real value since it is not possible from our data to estimate the inclination angle of each magnetic element. However, if we assume, hypothetically, that the magnetic elements are nearly vertical, then frequencies above $\sim 2.5-2.7$ mHz will be able to propagate upwards. This is an upper limit for the cutoff frequency itself since if the magnetic element is inclined to the vertical, the cutoff will be even smaller than what is estimated, allowing the propagation of oscillations at even smaller frequencies. Regardless, the fundamental mode at  $\sim 7.6 \pm 0.2$ min ($2.2$ mHz) is outside the propagation regime and not all the magnetic elements may have a sufficient inclination to lower the cutoff frequency enough to allow this mode to propagate. However, in small magnetic elements the cutoff frequency can also be reduced by radiative losses \citep{2008ApJ...676L..85K}. \\
For these reasons, we believe that the subharmonic response of the flux tubes to granular buffeting does not largely contribute to cromospheric heating. However the subharmonic response observed provides useful information on the mechanisms of excitation of the flux tubes and represents a strong observational proof of the interaction between the photospheric plasma and the magnetic elements themselves.\\
Besides this, it is worth noting that the persistence of the wave trains is larger at higher frequencies. This means that the amount of energy eventually contained in each wave train is larger and larger as the frequency increases. This is obviously an important point that has several implications for the heating of the upper layers of the Sun's atmosphere. Although our analysis is limited in frequency by the width of the kernel used to estimate the PDF, we can speculate that the measured increase of the persistence, seen up to $\nu \sim 1.6$ mHz ($T \sim 10$ min), can in fact extend to even higher frequencies that are allowed to propagate upward. In other words, even if the low-frequency kink oscillations were propagated they would have contributed little to the energy budget of the upper layers as their persistence is very short compared to their high frequency counterparts.

\section{Conclusions}
In this work we have analysed kink oscillations of several photospheric magnetic elements by employing the EMD technique. This technique is ideal for the study of non-stationary processes. This method allows us to decompose each velocity signal associated to the magnetic element in different modes using a decomposition basis which is derived empirically from the signal itself. By studying the ACF of each mode we have estimated the main periodicities and their relative PDF.\\
Our findings demonstrated the presence of numerous oscillation peaks which are consistent with subharmonic oscillations of a fundamental  whose period is $\sim 7.6 \pm 0.2$ min. Such a period is close to the characteristic temporal scale of the photospheric convection cells, therefore we can argue that these oscillations are associated to buffeting induced oscillations.\\

\begin{acknowledgements}
This work has been partly supported by the PRIN-INAF 2010 and PRIN-MIUR 2012 (prot. 2012P2HRCR) entitled "Il sole attivo ed i suoi effetti sul clima dello spazio e della terra" grants, funded by the Italian National Institute for Astrophysics (INAF) and Ministry of Education, Universities and Research (MIUR) respectively.\\
Hinode is a Japanese mission developed and launched by ISAS/JAXA, collaborating with NAOJ as a domestic partner, NASA and STFC (UK) as international partners. Scientific operation of the Hinode mission is conducted by the Hinode science team organized at ISAS/JAXA. This team mainly consists of scientists from institutes in the partner countries. Support for the post-launch operation is provided by JAXA and NAOJ (Japan), STFC (U.K.), NASA, ESA, and NSC (Norway). We also thank the anonymous referee for the valuable comments that helped a lot to improve the manuscript.\\
\end{acknowledgements}

\bibliographystyle{aa}

\end{document}